\documentclass[10pt,english]{article}
\usepackage{times}
\usepackage[T1]{fontenc}
\usepackage[latin1]{inputenc}
\usepackage{amsmath}
\usepackage{amssymb}

\makeatletter

\newcommand{\noun}[1]{\textsc{#1}}

\usepackage{slashed}

\usepackage{babel}
\makeatother
\begin{document}

\title{\textbf{Neutrino Mass Squared Differences in the Exact Solution of
a 3-3-1 Gauge Model without Exotic Electric Charges }}

\author{\noun{ADRIAN} PALCU}

\date{\emph{Department of Theoretical and Computational Physics - West
University of Timi\c{s}oara, V. P\^{a}rvan Ave. 4, RO - 300223 Romania}}

\maketitle
\begin{abstract}
The mass splittings for the Majorana neutrinos in the exact solution
of a particular 3-3-1 gauge model are computed here in detail. Since
both $\sin^{2}\theta_{13}\simeq0$ and the mass splittings ratio $r_{\Delta}\simeq0.033$
are taken into account, the analytical calculations seem to predict
an inverted mass hierarchy and a mixing matrix with a texture based
on a very close approximation to the bi-maximal mixing. The resulting
formulas for the mass squared differences can naturally accomodate
the available data if the unique free parameter ($a$) gets very small
values ($\sim10^{-15}$). Consequently, the smallness of the parameter
requires (according to our method) a large breaking scale $<\phi>\sim10^{6}-10^{7}$
TeV in the model. Hence, the results concerning the neutrino mass
splittings may lead to a more precise tuning in the exact solution
of the 3-3-1 model of interest, being able - at the same time - to
recover all the Standard Model phenomenology and predict the mass
spectrum of the new gauge bosons $Z^{\prime},X,Y$ in accordance with
the actual data. The minimal absolute mass in the neutrino sector
is also obtained - $m_{0}\simeq0.0035$ eV - in the case of our suitable
approximation for the bi-maxcimal mixing.


\end{abstract}

\section{Introduction}

The neutrino mass issue is one of the most effervescent still open
questions in the nowdays physics. Since the successful Standard Model
(SM) deals only with massless neutrinos that pair charged leptons
in three left-handed flavor generations of the gauge group $SU(3)_{c}\otimes SU(2)_{L}\otimes U(1)_{Y}$,
it appears obviously that any theoretical mechanism designed to generate
neutrino masses must invoke some new physics beyond the SM and even
a radically new framework. One such a natural extension of the SM
emerged in the literature with the papers of Pisano, Pleitez and Frampton
\cite{key-1} - based on the new gauge group $SU(3)_{c}\otimes SU(3)_{L}\otimes U(1)_{Y}$
- that opened a plethora of theoretical ways \cite{key-2} to generate
neutrino mases. These mathematical strategies have been developed
eversince in order to accomodate the available data \cite{key-3}
regarding the solar and atmospheric neutrino oscillation experiments,
but a final answer to this challenge has not been given yet. 

The author proposed an original mechanism \cite{key-4} in order to
provide neutrino masses using the exact solution of a particular class
of 3-3-1 models (namely, model D in \cite{key-5}). The neutrinos
are considered as Majorana fields and their mass matrix arises naturally
from the Lagrangian of the model when particular tensor products among
Higgs triplets are involved. This strategy works and relates the neutrino
masses to the charged lepton ones, since the SSB takes place in the
manner proposed by the general method of solving electro-weak gauge
models with high symmetries \cite{key-6}. It relies on a special
parametrization of the Higgs sector that finally provides a one-parameter
mass scale \cite{key-4}. The one-parameter solution of the 3-3-1
model under consideration here has the advantage that it recovers
all the SM phenomenology, it can predict the mass spectrum for the
new bosons in the model and it needs no massive Majorana seesaw partners
for physical neutrinos (even no seesaw mechanism at all). All these
results are achieved just by tunig the unique remaining free parameter
of the model. Besides, a large overall breaking scale seems to be
required due to the smallness of the free parameter $a$. Calculation
of the concrete expressions for the mass splittings within the one-parameter
exact solution of the model is accomplished here by taking into account
the condition $\sin^{2}\theta_{13}\simeq0$ which seems to be most
likely, since the actual experimental investigations are not sensitive
to any CP-phase violation in the lepton sector. 

The paper is organized as follows. In Sec. 2 the neutrino mass matrix
is obtained within the framework of the exact solution of the 3-3-1
model of interest. Then, the mass matrix is analytically diagonalized
by solving an appropriate set of equations (Sec. 3). Consequently,
the mass squared differences are computed and some phnomenological
implications are discussed in Sec.4, where a certain texture to the
mixing matrix is proposed. Our conclusions are sketched in the last
section of the paper (Sec. 5).

\section{Neutrino Mass Matrix in the 3-3-1 Model without Exotic Electric Charges}

In the following we look for the neutrino masses, recalling the main
results of the exact solution \cite{key-4} of a 3-3-1 model without
exotic electric charges (model D in \cite{key-5}) and assuming the
neutrino mixing in the manner presented in the excellent reviews of
Bilenky \cite{key-7}. All the standard notations of the field are
considered and used below. 

The neutrino mass splittings can be addressed only after defining
the physical neutrino mass eigenstates. A unitary mixing matrix $U$
(with $U^{+}U=1$) is necessary for this purpose. It links the gauge-flavor
basis to the physical one of the massive neutrinos: 

\begin{equation}
\nu_{\alpha L}(x)=\sum_{i=1}^{3}U_{\alpha i}\nu_{iL}(x)\label{Eq. 1}\end{equation}
where $\alpha=e,\mu,\nu$ (denoting gauge flavor-eigenstates) and
$i=1,2,3$ (denoting massive physical eigenstates). We consider hereafter
the physical neutrinos as Majorana fields, \emph{i.e.} $\nu_{iL}^{c}(x)=\nu_{iL}(x)$.
The neutrino mass term in the Yukawa sector yields then: 

\begin{equation}
\mathcal{-L}_{Y}=\frac{1}{2}\bar{\nu}_{L}M\nu_{L}^{c}+H.c\label{Eq. 2}\end{equation}
with $\nu_{L}=\left|\begin{array}{ccc}
\nu_{e} & \nu_{\mu} & \nu_{\tau}\end{array}\right|_{L}^{T}$ where the superscripts $T$ and $c$ mean ''transposed'' and ''charge
conjugation'' respectively, while subscript $L$ means ''left-handed''
component. The mixing matrix $U$ that diagonalizes the mass matrix
in the manner $U^{+}MU=m_{ij}\delta_{j}$ has in the standard parametrization
the form: 

\begin{equation}
U=\left|\begin{array}{ccc}
c_{2}c_{3} & s_{2}c_{3} & s_{3}e^{-i\delta}\\
-s_{2}c_{1}-c_{2}s_{1}s_{3}e^{i\delta} & c_{1}c_{2}-s_{2}s_{3}s_{1}e^{i\delta} & c_{3}s_{1}\\
s_{2}s_{1}-c_{2}c_{1}s_{3}e^{i\delta} & -s_{1}c_{2}-s_{2}s_{3}c_{1}e^{i\delta} & c_{3}c_{1}\end{array}\right|\label{Eq. 3}\end{equation}
where we have denoted $\sin\theta_{23}=s_{1}$, $\sin\theta_{12}=s_{2}$,
$\sin\theta_{13}=s_{3}$, $\cos\theta_{23}=c_{1}$, $\cos\theta_{12}=c_{2}$,
$\cos\theta_{13}=c_{3}$ for the mixing angles, and $\delta$ for
the CP phase. 

In order to generate the expected small Majorana masses for neutrinos,
Ref. \cite{key-4} proposed (within the exact solution of the 3-3-1
model D) an original and eficient method based on a new term in the
Yukawa sector, namely: $G_{\alpha\beta}{}_{L}\bar{f}_{\alpha L}Sf_{\beta L}^{c}+H.c.$,
with $S\sim(\mathbf{1},\mathbf{6},-2/3)$ and $G_{\alpha\beta}$ as
the coupling coefficients of the lepton sector. The proposed symmetric
matrix $S$ suitable for our purpose is constructed as $S=\phi^{-1}\left(\phi^{(\eta)}\otimes\phi^{(\chi)}+\phi^{(\chi)}\otimes\phi^{(\eta)}\right)$. 

We recall that the exact solution of the 3-3-1 model of interest leads
to the one-parameter matrix $\eta=\left(1-\eta_{0}^{2}\right)diag\left[a/2\cos^{2}\theta_{W},1-a,a\left(1-\tan^{2}\theta_{W}\right)/2\right]$
which determines the Higgs sector $\phi^{(i)}=\eta^{i}\phi,i=1,2,3$
and the VEV alignment (successively the SSB) \cite{key-4} in according
to the general prescriptions of the method shown in Ref. \cite{key-6}. 

Our procedure provides \cite{key-4} the following symmetric mass
matrix for the neutrinos involved in the model:

\begin{equation}
M=4\left|\begin{array}{ccc}
A & D & E\\
D & B & F\\
E & F & C\end{array}\right|\frac{<\phi^{(\eta)}><\phi^{(\chi)}>}{<\phi>}\label{Eq. 4}\end{equation}
while the charged leptons of the model acquire the masses $m(e)=A<\phi^{(\rho)}>$,
$m(\mu)=B<\phi^{(\rho)}>$, $m(\tau)=C<\phi^{(\rho)}>$. Obviously,
$G_{ee}=A$, $G_{\mu\mu}=B$, $G_{\tau\tau}=C$, $G_{e\mu}=D$, $G_{e\tau}=E$,
$G_{\mu\tau}=F$. 

Based on the trace properties in (4), the mass ratio (neutrinos to
charged leptons) yields: 

\begin{equation}
\frac{\sum_{i}m_{i}}{m(e)+m(\mu)+m(\tau)}=4\frac{<\phi^{(\eta)}><\phi^{(\chi)}>}{<\phi^{(\rho)}><\phi>}\label{Eq. 5}\end{equation}
where the left hand term of Eq. (5) is phenomenologically restricted
by the actual data \cite{key-8}, while the right hand term can be
tuned by means of certain theoretical computations. The latter is
determined by the parameter choice (namely, the bijective mapping
$(\eta,\rho,\chi)\rightarrow(1,2,3)$) that establish a certain VEV
alignment. Therefore, in order to accomodate the data concerning the
sum of the neutrino masses, one can accept only the Case I (see for
details Sec. 4.3 in \cite{key-4}) for which the parameter $a$ (that
determines the VEV alignment) has to be in the range $a<0.118\cdot10^{-9}$\cite{key-4}.
It follows naturally that $<\phi_{1}>,<\phi_{3}>\ll<\phi_{2}>$. Under
these circumstances, the mass matrix becomes: 

\begin{equation}
M=\left|\begin{array}{ccc}
m(e) & D & E\\
D & m(\mu) & F\\
E & F & m(\tau)\end{array}\right|\left(\frac{2a}{\sqrt{1-a}}\right)\frac{\sqrt{1-2\sin^{2}\theta_{W}}}{\cos^{2}\theta_{W}}\label{Eq. 6}\end{equation}
where - rigorously speaking - the new $D,E,F$ in Eq. (6) differ from
those in Eq. (4) by a factor $<\phi^{(\rho)}>$. However this is not
a reason to change the notations since they are amounts to be eliminated
afterwards in our calculations.

\section{Calculating Neutrino Masses }

At the present level of knowledge, the absolute masses of the physical
neutrinos are experimentally irrelevant. Instead, the mass squared
differences - defined as $\Delta m_{ij}^{2}=m_{j}^{2}-m_{i}^{2}$
- can offer significant data by observing the neutrino oscillations
phenomenon. Their right order of magnitude is $7.1\times10^{-5}$
eV$^{2}$ $\leq\Delta m_{12}^{2}\leq$ $8.9\times10^{-5}$ eV$^{2}$
(from solar and KamLAND data \cite{key-3}) and $1.4\times10^{-3}$
eV$^{2}$ $\leq\Delta m_{13}^{2}\leq$ $3.3\times10^{-3}$ eV$^{2}$
(from Super Kamiokande atmospheric data \cite{key-3}). A systematic
approach of the way to deal with them can be found in \cite{key-10}
and references therein. These data are compatible with either a normal
mass hierarchy $\Delta m_{23}^{2}>0$ ($m_{1}<m_{2}<m_{3}$) or an
inverted one $\Delta m_{23}^{2}<0$ ($m_{3}\ll m_{1}\simeq m_{2}$). 

For our purpose, the coupling constants in Eq. (4) are regarded as
unknown variables and try to eliminate them by diagonalization the
matrix $M$. This is equivalent (in the Majorana case) to a set of
6 linear equations with 9 variables which can supply the following
generic solution for the physical neutrino masses:

\begin{equation}
m_{i}=f_{i}\left[m(e),m(\mu),m(\tau)\right]\left(\frac{2a}{\sqrt{1-a}}\right)\frac{\sqrt{1-2\sin^{2}\theta_{W}}}{\cos^{2}\theta_{W}}\label{Eq. 7}\end{equation}
with $i=1,2,3$. 

The concrete forms of $f_{i}s$ remain to be determined by solving
the following set of equations:

\begin{equation}
\begin{cases}
\begin{array}{c}
m_{1}=c_{2}^{2}m(e)+c_{1}^{2}s_{2}^{2}m(\mu)+s_{1}^{2}s_{2}^{2}m(\tau)-2c_{1}c_{2}s_{2}D+2s_{1}s_{2}c_{2}E-2c_{1}s_{1}s_{2}^{2}F\\
0=c_{2}s_{2}m(e)-c_{1}^{2}c_{2}s_{2}m(\mu)-s_{1}^{2}s_{2}c_{2}m(\tau)-(1-2s_{2}^{2})s_{1}E+2s_{1}s_{2}c_{1}c_{2}F\\
0=-c_{1}s_{1}s_{2}m(\mu)+c_{1}s_{1}s_{2}m(\tau)+c_{2}s_{1}D+c_{1}c_{2}E-(1-2s_{1}^{2})s_{2}F\\
m_{2}=s_{2}^{2}m(e)+c_{1}^{2}c_{2}^{2}m(\mu)+s_{1}^{2}c_{2}^{2}m(\tau)+2c_{1}c_{2}s_{2}D-2s_{1}s_{2}c_{2}E-2c_{1}s_{1}c_{2}^{2}F\\
0=s_{1}c_{1}c_{2}m(\mu)-s_{1}c_{1}c_{2}m(\tau)+s_{1}s_{2}D+c_{1}s_{2}E+(1-2s_{1}^{2})c_{2}F\\
m_{3}=s_{1}^{2}m(\mu)+c_{1}^{2}m(\tau)+2c_{1}s_{1}F\end{array}\end{cases}\label{Eq. 8}\end{equation}

Since actual data are not sensitive to any CP-phase violation in the
lepton sector we have taken into accound from the very beginning $\sin^{2}\theta_{13}\simeq0$
- as it can be easily observed by inspecting the shape of Eq. (8)
- but the proposed values for the other two mixing angles will be
embedded only in the resulting formulas for the neutrino masses (9).
Thus, one obtaines after a few manipulations the following analytical
equations:

\[
m_{1}=\frac{m(\tau)\sin^{2}\theta_{12}\sin^{2}\theta_{23}-m(\mu)\sin^{2}\theta_{12}\left(1+\sin^{2}\theta_{23}\right)}{\left(1-2\sin^{2}\theta_{23}\right)\left(1-2\sin^{2}\theta_{12}\right)}+\frac{m(e)\left(1-\sin^{2}\theta_{12}\right)}{\left(1-2\sin^{2}\theta_{12}\right)},\]

\[
m_{2}=\frac{m(\mu)(1-\sin^{2}\theta_{12}-\sin^{2}\theta_{12}+3\sin^{2}\theta_{12}\sin^{2}\theta_{23})-m(\tau)\sin^{2}\theta_{23}\left(1-\sin^{2}\theta_{12}\right)}{\left(1-2\sin^{2}\theta_{23}\right)\left(1-2\sin^{2}\theta_{12}\right)}\]

\[
+\frac{m(e)\sin^{2}\theta_{12}}{\left(1-2\sin^{2}\theta_{12}\right)},\]

\begin{equation}
m_{3}=\frac{m(\tau)\left(1-\sin^{2}\theta_{23}\right)-m(\mu)\sin^{2}\theta_{23}}{1-2\sin^{2}\theta_{23}}.\label{Eq. 9}\end{equation}
A rapid investigation of the resulting equations can be made even
at this stage. 

\begin{itemize}
\item Note that some of the masses could come out negative (for certain
combinations of angles), but this is not an impediment since for any
fermion field a $\gamma_{5}\psi$ transformation can be performed
at any time, without altering the physical content of the theory.
As a result of this manipulation the mass sign changes. 
\item Let's observe that the analytical euqtions for the masses (9) strictly
impose $\sin^{2}\theta_{12}\neq0.5$ and $\sin^{2}\theta_{12}\neq0.5$
, yet this does not forbid any closer approximation to the bi-maximal
neutrino mixing. 
\item These equations also do not contradict the trace condition which requires
indeed a finite neutrino mass sum independently of the values of the
mixing angles. As a matter of fact, if one summes the three masses
in (9), then the troublesome denominators disappear and the required
(by Eq. (6)) value is recovered. 
\item The particular shape of the analytical neutrino masses is due to both
the choice of the $\theta_{13}=0$ and the diagonal entries in the
mixing matrix that are proportional to the charged lepton masses.
Any other choice definitely leads to a different set of equations
to be solved and, thus, to a different form of the solution.
\item The mass splittings ratio is independent of the free parameter of
the model, therefore its right value can be determined only by chosing
appropriate values for the mixing angles.
\item Assuming the structure of the mass spectrum in the charged lepton
sector \cite{key-9}, one can neglect in Eqs. (9) all the terms except
for those multipling $m(\tau)$. That is, our analysis will rely on
the following suitable approximation:
\end{itemize}
\[
m_{1}\simeq\frac{m(\tau)\sin^{2}\theta_{12}\sin^{2}\theta_{23}}{\left(1-2\sin^{2}\theta_{23}\right)\left(1-2\sin^{2}\theta_{12}\right)}\]

\begin{equation}
\left|m_{2}\right|\simeq\frac{m(\tau)\left(1-\sin^{2}\theta_{12}\right)\sin^{2}\theta_{23}}{\left(1-2\sin^{2}\theta_{23}\right)\left(1-2\sin^{2}\theta_{12}\right)}\label{Eq. 10}\end{equation}

\[
m_{3}\simeq\frac{m(\tau)\left(1-2\sin^{2}\theta_{12}\right)\left(1-\sin^{2}\theta_{23}\right)}{\left(1-2\sin^{2}\theta_{23}\right)\left(1-2\sin^{2}\theta_{12}\right)}\]

\section{Phenomenological Implications}

One of the striking features of the above presented solution is that
it requires in the most likely case (see below) an \textbf{inverted
hierarchy} $m_{3}\ll m_{1}\simeq m_{2}$, since the data \cite{key-10}
suggest the following limits for the solar mixing angle: $(1/3)\leq\sin^{2}\theta_{12}\leq(1/2)$.
Hence, $m_{1}\leq\left|m_{2}\right|$holds in (10) as long as $\sin^{2}\theta_{12}\leq1/2$.
If the maximal atmospheric angle is embedded, then $m_{3}<m_{1}$
is accomplished in (10) only if $\sin^{2}\theta_{12}>1/3$ which is
quite a plausible prediction.

\subsection{Case 1 ($\sin^{2}\theta_{23}\simeq0.5$)}

We have already established that the most likely setting with maximal
atmospheric angle occurs within the inverted hierarchy. In the following
we exploit the mass splittings ratio $r_{\Delta}=\Delta m_{12}^{2}/\Delta m_{23}^{2}$
that has to be fulfilled ($r_{\Delta}\simeq0.033$) by a certain value
of the solar mixing angle. A suitable approximation will be performed
in the denominators - which are identical for all the three terms
in Eqs. (10) - only as the last step of the calculation, while for
the moment we focus on the numerators which determine the splitting
ratio. They are:

\[
m_{1}\sim m(\tau)\sin^{2}\theta_{12}\]

\begin{equation}
\left|m_{2}\right|\sim m(\tau)\left(1-\sin^{2}\theta_{12}\right)\label{Eq. 11}\end{equation}

\[
m_{3}\sim m(\tau)\left(1-2\sin^{2}\theta_{12}\right)\]

Under these circumstances - \emph{i.e}. inverted hierarchy - the mass
ratio will have the form:

\begin{equation}
r_{\Delta}=\frac{\left(1-2\sin^{2}\theta_{12}\right)}{\left(2-3\sin^{2}\theta_{12}\right)\sin^{2}\theta_{12}}\label{Eq. 12}\end{equation}

Since one takes as a good approximation in the numerator of Eq. (12)
$\sin^{2}\theta_{12}\simeq0.495$, a reasonable value for $r_{\Delta}$can
result very close to the phenomenological value $0.033$. (Actually,
the value for $\sin^{2}\theta_{12}$was inferred by solving the resulting
equation (12) with the solar mixing angle as variable; a unique solution
out of the two is suitable, while the other one is rejcted not belonging
to $[-1,1]$)

Once the solar mixing angle has been established, the mass squared
differences can be computed. The solar neutrino mass splitting is:

\begin{equation}
\Delta m_{12}^{2}=\frac{m^{2}(\tau)}{\left(1-2\sin^{2}\theta_{12}\right)\left(1-2\sin^{2}\theta_{23}\right)^{2}}\left(\frac{a^{2}}{1-a}\right)\frac{1-2\sin^{2}\theta_{W}}{16\cos^{4}\theta_{W}}\sim10^{25}a^{2}eV^{2}\label{Eq. 13}\end{equation}
According to the data supplied by \cite{key-9} for the Weinberg angle
and tau lepton and assuming the approximation $\sin^{2}\theta_{23}=0.4995$,
the resulting value for parameter $a$ has to be $\sim2\times10^{-15}$
in order to fit the experimental condition $\Delta m_{12}^{2}\simeq8\times10^{-5}$eV$^{2}$.
Consequently, this value of the parameter fits the atmospheric neutrino
mass splitting$\Delta m_{23}^{2}\simeq2.4\times10^{-3}$ eV$^{2}$
as well. Such a small value for the free parameter determines a large
VEV $<\phi>\sim10^{6}-10^{7}$TeV.

Thus, the bi-maximal condition seems to successfully accomodate our
method. Moreover, it can predict even the minimal absolute mass in
the neutrino sector. It is now expressed by:

\begin{equation}
m_{3}\simeq\frac{m(\tau)}{\left(1-2\sin^{2}\theta_{23}\right)}\left(\frac{a}{\sqrt{1-a}}\right)\frac{\sqrt{1-2\sin^{2}\theta_{W}}}{\cos^{2}\theta_{W}}\label{Eq. 14}\end{equation}
and has the approximate value $m_{3}\simeq0.0035$ eV.

\subsection{Case 2 ($\sin^{2}\theta_{12}\simeq0.3$)}

Let's have a look at this particular case. The question is whether
the lower limit $\sin^{2}\theta_{12}\simeq0.3$ can supply a suitable
setting in our method. A rapid estimation in this case confirms also
an inverted herarchy where the numerators of the individual neutrino
masses can be sufficiently well approximated as:

\[
m_{1}\sim0.3m(\tau)\sin^{2}\theta_{23}\]

\begin{equation}
\left|m_{2}\right|\sim0.7m(\tau)\sin^{2}\theta_{23}\label{Eq. 15}\end{equation}

\[
m_{3}\sim0.4m(\tau)\left(1-\sin^{2}\theta_{23}\right)\]

A normal hierarchy would require - when comparing $m_{3}>m_{2}$ -
an unacceptable upper limit for the atmospheric angle $\sin^{2}\theta_{23}<4/11\sim0.36$.
Thus, $\sin^{2}\theta_{23}>4/7\sim0.57$ which ensures $m_{3}<m_{1}$
seems to be the unique acceptable possibility. Unfortunately, this
leads to a similar equation to (12):\begin{equation}
r_{\Delta}=\frac{4\sin^{4}\theta_{23}}{3.3\sin^{4}\theta_{23}+3.2\sin^{2}\theta_{23}-1.6}\label{Eq. 16}\end{equation}
that gives unacceptable solutions for $\sin^{2}\theta_{23}$(both
resulting values not belonging to $[-1,1]$) when $r_{\Delta}=0.033$
is embedded in (16). Therefore, this case is ruled out.

\section{Concluding Remarks}

In conclusion, we consider that we have convincingly proved that our
method of exactly solving gauge models with high symmetries \cite{key-4,key-6}
when applied to the particular 3-3-1 model without exotic electric
charges can provide a realistic solution \cite{key-4}. That is, it
can recover all the SM phenomenology and predict a plausible mass
spectrum for the new bosons $Z^{\prime},X,Y$ (their masses are $\geq$
$1$ TeV, in good accordance to \cite{key-9}). Furthermore, if a
special Yukawa mechanism based on tensor products among Higgs triplets
is implemented into the model, then neutrino masses can be generated
in a natural way at the tree level without any approximation \cite{key-4}.
The resulting values can accomodate the available data concerning
the mass splittin \cite{key-8,key-9,key-10} just by tuning a unique
free parameter ($a$) as we have just shown in the previous section,
even though this requires a large breaking scale for one of the three
VEVs $<\phi>\sim10^{6}-10^{7}$TeV (the overall scale). If one wants
to separate the VEV question from the neutrino mass issue (and hence,
getting lighter non-SM bosons), one has to add a new small parameter
in a proper way in the parameter matrix of the Higgs sector. This
mathematical artifice \cite{key-11} will generate a canonical seesaw
mechanism that can be naturally implemented in the model even at a
low breaking scale $\sim1$TeV (the procedure to be presented elsewhere). 

Concerning the texture of the mixing matrix obtained above (Sec. 4.1),
note that this kind of settings were widely discused in the literature
\cite{key-12,key-13,key-14}. It implies - along with the inverted
hierarchy - the approximate symmetry $L_{e}-L_{\mu}-L_{\tau}$ \cite{key-13}.
Although some predictions of this scenario (one of them: the maximal
solar mixing requirement, in the leading order of perturbations) are
not in vogue, analyzing the topic is worthwhile (see \cite{key-14}
and references therein) since the issue of the neutrino matrix texture
awaits more accurate evidence in the forthcoming experiments.

Our method definitely favors the bi-maximal mixing and inverted hierarchy
for physical neutrinos.

\end{document}